# Quantum Paraelectricity in the Kitaev Quantum-Spin-Liquid Candidates $H_3LiIr_2O_6$ and $D_3LiIr_2O_6$

K. Geirhos,[1] P. Lunkenheimer,[1,*] M. Blankenhorn,[2] R. Claus,[2] Y. Matsumoto,[2] K. Kitagawa,[3] T. Takayama,[2,4] H. Takagi,[2,4] I. Kézsmárki,[1] and A. Loidl[1]

[1] *Experimental Physics V, Center for Electronic Correlations and Magnetism, University of Augsburg, 86159 Augsburg, Germany*
[2] *Max Planck Institute for Solid State Research, 70569 Stuttgart, Germany*
[3] *Department of Physics, University of Tokyo, Tokyo 113-0033, Japan*
[4] *Institute for Functional Matter and Quantum Technologies, University of Stuttgart, 70569 Stuttgart, Germany*

$H_3LiIr_2O_6$ is the first honeycomb-lattice system without any signs of long-range magnetic order down to the lowest temperatures, raising the hope for the realization of an ideal Kitaev quantum spin liquid. Its honeycomb layers are coupled by interlayer hydrogen bonds. Static or dynamic disorder of these hydrogen bonds was proposed to strongly affect the magnetic exchange and to make Kitaev-type interactions dominant. Using dielectric spectroscopy, here we provide experimental evidence for dipolar relaxations in $H_3LiIr_2O_6$ and deuterated $D_3LiIr_2O_6$, which mirror the dynamics of protons and deuterons within the double-well potentials of the hydrogen bonds. The detected hydrogen dynamics reveals glassy freezing, characterized by a strong slowing down under cooling, with a crossover from thermally-activated hopping to quantum-mechanical tunneling towards low temperatures. Thus, besides being Kitaev quantum-spin-liquid candidates, these materials also are quantum paraelectrics. However, the small relaxation rates in the mHz range, found at low temperatures, practically realize quasi-static hydrogen disorder, as assumed in recent theoretical works to explain the quantum-spin-liquid ground state of both compounds.

## I. INTRODUCTION

When a magnetic material with interacting spins $S = ½$ approaches zero temperature without undergoing long-range magnetic ordering or any kind of spin freezing, this exotic state of matter is called a quantum spin liquid (QSL) [1,2]. Since decades, QSLs are in the focus of solid-state research and, in most cases, the dimensionality of the spin system or underlying frustration effects are the reason for the suppression of long-range spin order. In 2006, Kitaev [3] proposed an exactly solvable model of frustrated quantum spins $S = ½$ on a two-dimensional honeycomb lattice with bond-directional interactions, in which the spins fractionalize into Majorana fermions and form a topological QSL. Somewhat later, Jackeli and Khaliullin [4] demonstrated that in Mott insulators with strong spin-orbit coupling and with a $J_{eff} = ½$ ground state, these bond-directional interactions indeed could be realized and iridium oxides were identified as possible Kitaev-type spin-liquid candidates. Since then, a number of 4d and 5d materials were considered to exhibit a possible QSL ground state, with $Na_2IrO_3$ [5], $\alpha$-$Li_2IrO_3$ [6], and $\alpha$-$RuCl_3$ [7] being the most prominent examples. However, despite the importance of Kitaev-type interactions in all of these compounds, their ground states always reveal long-range magnetic order. This documents the relevance of further interactions, like the isotropic Heisenberg or off-diagonal exchange terms, bringing the systems away from the pure Kitaev limit. For a review on concepts and material realizations of Kitaev spin liquids see Refs. [8,9].

Very recently, Takagi and colleagues [10] have synthesized a new QSL candidate material, $H_3LiIr_2O_6$, by hydrogen substitution of the interlayer lithium in $\alpha$-$Li_2IrO_3$. In this compound all the interlayer $Li^+$ ions are replaced by $H^+$, leaving the $LiIr_2O_6$ honeycomb plane intact [11]. In this material all honeycomb layers are coupled via interlayer O–H···O hydrogen bonds as schematically indicated in Fig. 1. Interestingly, this material shows no signs of magnetic order down to 50 mK, despite of magnetic interactions of the order of 100 K [10], raising the hope for the realization of an ideal Kitaev QSL. Recent Raman-scattering results, detecting a magnetic continuum, give further support for a Kitaev QSL in $H_3LiIr_2O_6$ [12]. Initial studies of the deuterated isostructural variant $D_3LiIr_2O_6$ also suggest QSL behavior [13].

Obviously, in this material the Kitaev exchange $K$ exceeds other interactions, like the Heisenberg coupling $J$ or the off-diagonal exchange $\Gamma$, that lead to magnetic order in the related iridate systems. Subsequent to the pioneering work of Ref. [10], different mechanisms for a dominating Kitaev interaction in $H_3LiIr_2O_6$ were proposed in literature, essentially based on two alternative types of disorder: (i) Various forms of hydrogen disorder, both static [14,15] or dynamic [16,17]. (ii) Stacking faults within the succession of the $LiIr_2O_6$ honeycomb layers [18,19], for which $H_3LiIr_2O_6$ is known to be prone [11]. Here one should be aware that the layer stacking also influences the hydrogen-bond geometry [11,19], which in turn affects the Kitaev-type interactions [19]. Within both scenarios, not only the dominance of the Kitaev exchange leading to a QSL could be explained, but also the specific-heat and NMR spin-relaxation results from Ref. [10], which at first glance seemed at odds with the Kitaev model [15,18,20]. Finally, one should be aware that defect-induced disorder without invoking any Kitaev exchange may also explain the observed spin-liquid state, e.g., simply via a disorder-induced reduction of $J$ or by frustration effects [16,21]. Such defects may be, e.g., hydrogen vacancies or an imperfect Li-H substitution arising during the synthesis of $H_3LiIr_2O_6$ from $\alpha$-$Li_2IrO_3$.

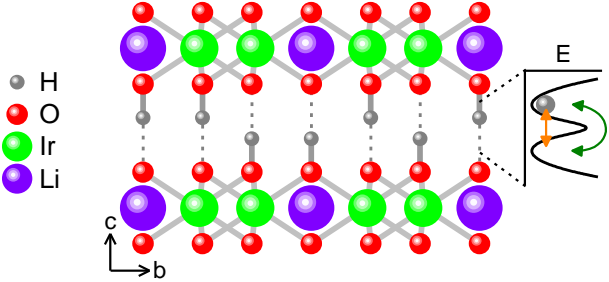

FIG. 1. Schematic structure of $H_3LiIr_2O_6$ showing two honeycomb layers [16]. A disordered arrangement of the interlayer hydrogen atoms is indicated as suggested by the results of the present work. For the hydrogens, the solid and dashed lines denote OH-bonds with the hydrogens statistically located at one side of the double-well potential. At the right side of the figure, the double-well potential experienced by a hydrogen atom is schematically indicated; the arrows symbolize thermally-activated hopping (prevailing at high temperatures) and quantum-mechanical tunneling (at low temperatures).

To check for the hydrogen-disorder scenarios mentioned above, dielectric spectroscopy is the prime experimental choice as was also pointed out in Ref. [17]. As indicated in Fig. 1, the protons within hydrogen bonds often experience a double-well potential and, indeed, this also is the case for $H_3LiIr_2O_6$ [16]. The resulting proton hopping between the potential wells corresponds to the fluctuation of a large dipole moment, which can be detected by dielectric spectroscopy. The ability of this method to directly monitor the proton dynamics in hydrogen bonds was documented in detail, e.g., for mixed molecular systems like (Rb:NH$_4$) dihydrogen phosphate [22] or for solid solutions of betaine phosphate and phosphite [23]. Höchli *et al.* have summarized in detail the freezing of dipolar moments in the solid state [24].

The time scale of the proton-hopping dynamics, which can be detected by dielectric spectroscopy, is of utmost importance for the magnetic coupling. The hydrogen disorder can be expected to be of dynamical nature at high temperatures, governed by thermal activation. It may freeze in, either randomly or in an ordered structure, under cooling. Due to the low proton mass, quantum-mechanical tunneling can play a role at low temperatures [16,17]. For sufficiently slow hydrogen dynamics, the resulting disorder may be regarded as quasi static, which corresponds to the scenarios considered in Refs. [14,15], increasing the relevance of the Kitaev exchange. Alternatively, in a recent theoretical work [16], it has been proposed that due to quantum fluctuations the hydrogen bonds may move quickly. Then, on time average the O–H–O bonds are symmetric, which was shown to enhance the Kitaev interaction as well [16]. In Ref. [17], dipolar fluctuations also were assumed to renormalize the magnetic couplings and help establishing a Kitaev QSL state.

$H_3LiIr_2O_6$ can be fully deuterated. In deuterated compounds, thermally activated behavior should essentially remain unchanged, while there should be significant effects in the tunneling regime. It has been proposed that upon deuteration the hydrogen bonds become asymmetric on time average [16], which should modify the magnetic coupling.

To unravel the hydrogen-bond dynamics of the title compounds, in the present work we investigate the temperature dependence of the dipolar dynamics utilizing dielectric spectroscopy applied to single-crystalline and polycrystalline $H_3LiIr_2O_6$, as well as polycrystalline $D_3LiIr_2O_6$. We experimentally pin down the proton and deuteron dynamics and document a transition from thermally activated behavior to a tunneling regime under cooling, leading to a quantum-paraelectric state. The high-temperature relaxations follow a super-Arrhenius behavior, indicative for glassy freezing. Our results reveal quasi-static hydrogen-bond disorder at low temperatures in both materials.

## II. EXPERIMENTAL DETAILS

$H_3LiIr_2O_6$ in polycrystalline form or as small single crystals were synthesized as described earlier [10,11]. The similar magnetic behavior of the single- and polycrystalline material is documented in Appendix A. Deuterated isostructural compounds were synthesized in polycrystalline form [13]. To obtain high-resolution dielectric measurements down to liquid He temperature and for frequencies $10^{-4} \leq \nu \leq 10^6$ Hz, a frequency-response analyzer (Novocontrol Alpha-A analyzer) was used [25]. The samples were prepared as parallel-plate capacitors. The single crystal had a surface area of about 0.024 mm$^2$ and a thickness of about 0.019 mm. The ceramic compounds were prepared as pressed pellets, with dimensions of 0.27 mm$^2$ × 0.19 mm for the protonated and of 2.32 mm$^2$ × 0.52 mm for the deuterated material. Electrical contacts were applied using silver paint. For sample cooling, a $^4$He-bath cryostat (Cryovac) and a closed-cycle refrigerator (Janis) were utilized.

## III. RESULTS AND DISCUSSION

Figures 2(a) and (b) show the temperature dependences of the real parts of the dielectric constant $\varepsilon'$ and the conductivity $\sigma'$, respectively, for various measuring frequencies as determined in single-crystalline $H_3LiIr_2O_6$. The single crystals used in these experiments were very small and irregularly shaped platelets. Hence, the absolute values in Fig. 2 only can provide the correct order of magnitude. At low temperatures and low frequencies, $\varepsilon'(T)$ exhibits a well-defined steplike decrease from a plateau regime of $\varepsilon' \approx 30$ to a low-temperature value of $\approx 8$. Such a step in the dielectric constant, which shifts to higher temperatures on increasing frequency, is the fingerprint of a relaxation process, signifying dipolar dynamics that slows down with decreasing temperature [26,27,28]. The only dipolar relaxation possible in $H_3LiIr_2O_6$ is the proton hopping within the double-well potential of the hydrogen bonds, which obviously is readily detectable by dielectric spectroscopy as discussed above. The step height in $\varepsilon'(T)$ represents the dipolar strength of this process and the temperature-independent plateau value in Fig. 2 signifies an almost temperature-invariant static dielectric constant. The steep increase of the dielectric constant following the steplike increase towards high temperatures is a non-intrinsic effect and due to electrode polarization [29].

Figure 3 shows the temperature-dependent dielectric results as obtained in polycrystalline samples. Here we only focus on the real part of the dielectric constant and compare measurements on protonated [Fig. 3(a)] and deuterated ceramics [Fig. 3(b)]. Just as for the single crystal [Fig. 2(a)], the dielectric constant in deuterated and protonated ceramic samples exhibits a steplike decrease under cooling. Again, its shift to higher temperatures on increasing frequency is a signature of dipolar relaxation dynamics that slows down on decreasing temperatures [26,28].

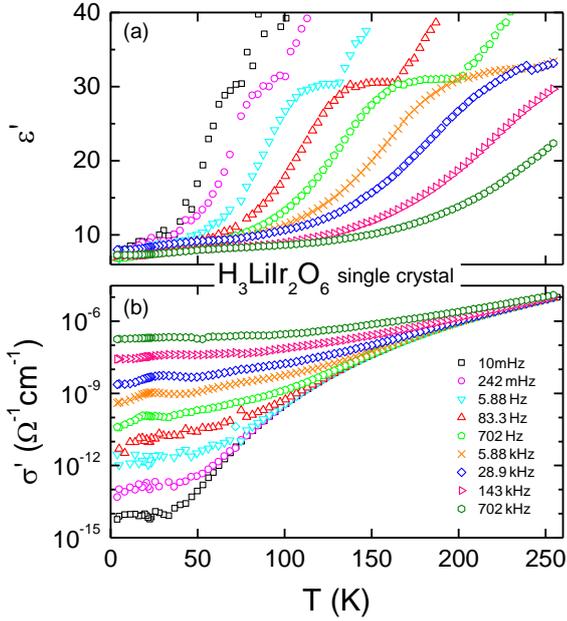

FIG. 2. Temperature dependence of (a) the real part of the dielectric constant $\varepsilon'$ and (b) the real part of the conductivity $\sigma'$ as measured in single-crystalline $H_3LiIr_2O_6$ shown for a series of frequencies.

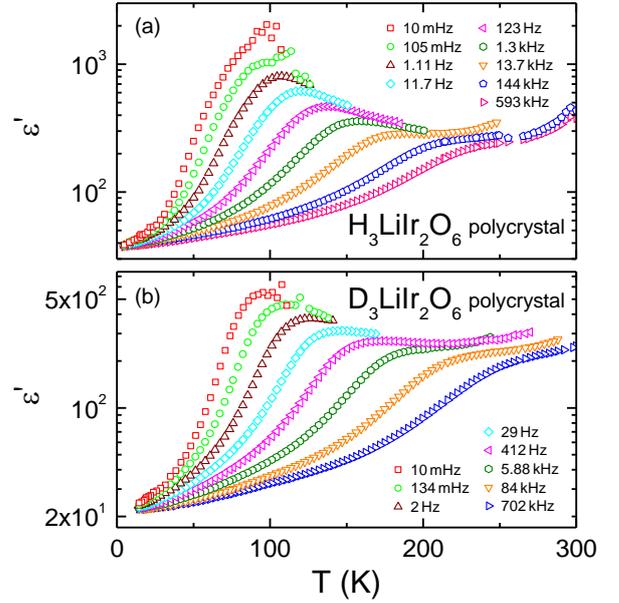

FIG. 3. Temperature dependence of the real part of the dielectric constant $\varepsilon'$ in (a) polycrystalline $H_3LiIr_2O_6$ and (b) polycrystalline $D_3LiIr_2O_6$ shown for a series of measuring frequencies.

The conductivity in Fig. 2(b) clearly exhibits semiconducting characteristics at all frequencies and temperatures. The 10 mHz results between 50 and 250 K can be interpreted as pure dc conductivity, $\sigma_{dc}$. The resulting dc resistivities $\rho_{dc} = 1/\sigma_{dc}$ for all compounds investigated are shown in Appendix B. Their temperature dependence follows Mott's variable-range-hopping law [30,31,32], speaking for the relevance of disorder in these samples. Notably, the results are in good agreement with the dc resistivity of ceramic samples as documented for a smaller temperature range in the Extended Data of Ref. [10]. The relaxation process detected in $\varepsilon'(T)$ [Fig. 2(a)] should lead to a peak in the dielectric loss $\varepsilon''$ and, thus, to a peak in $\sigma'(T)$ which is proportional to $\varepsilon''(T)\nu$ [26,28]. However, in Fig. 2(b) it is superimposed by the dc conductivity and only partly revealed by the crossover to a weaker temperature dependence of $\sigma'(T)$ [compared to $\sigma_{dc}(T)$] at low temperatures.

As documented in Fig. 3, the static dielectric constant of these ceramic samples reveals an increase towards low temperatures. This contrasts the behavior as observed in the single crystal, where the plateau region of $\varepsilon'(T)$ is rather temperature independent [Fig. 2(a)]. A Curie or Curie-Weiss like increase of the dielectric constant with decreasing temperatures is the typical temperature dependence of independent or weakly coupled dipoles, signaling a divergence of the dipolar susceptibility at zero or finite temperatures, respectively. It should be noted that the pattern of temperature- and frequency-dependent dielectric constants, as observed in Fig. 3, reminds of the typical signatures of relaxor ferroelectricity [27,33,34]. In relaxor ferroelectrics, dipolar regions of polar micro- or nano-domains cooperatively freeze-in, but do not establish true long-range ferroelectric order [35]. Thus, in these ceramic iridate samples the dipoles, which in the present case are related to the proton/deuteron positions within the hydrogen bonds, may exhibit ferroelectric-like short-range correlations. The different temperature dependences of the static dielectric constant in single-crystalline and ceramic samples probably signal a sample dependence of the evolution of dipolar interactions. Possibly, a higher density of stacking faults suppresses the development of three-dimensional ferroelectric order in the single crystal. However, as we will see in the following, the mean dipolar relaxation times in poly- and single crystals agree astonishingly well.

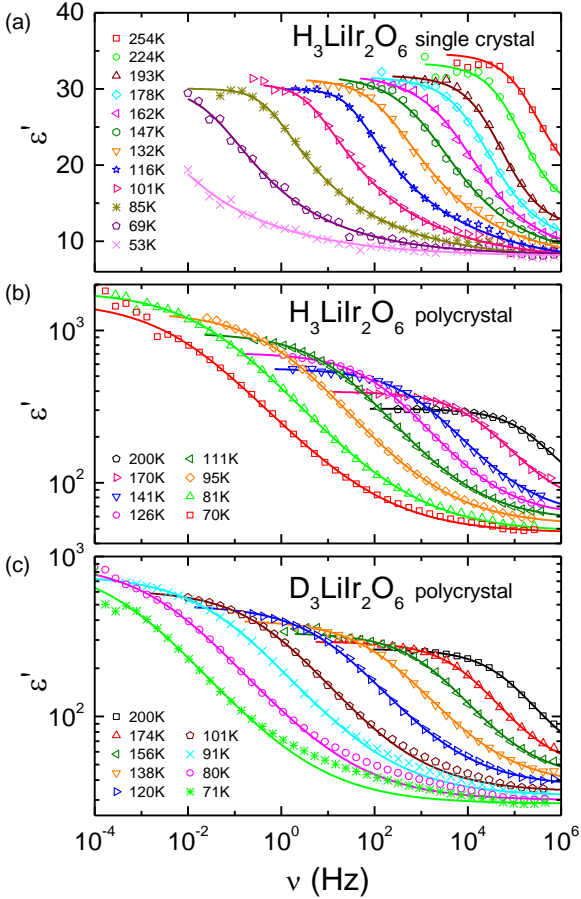

FIG. 4. Frequency dependence of the dielectric constant $\varepsilon'$ for (a) single crystalline and (b) polycrystalline $H_3LiIr_2O_6$, as well as for (c) deuterated polycrystalline $D_3LiIr_2O_6$. Data within the frequency region dominated by the observed relaxation processes are shown for selected temperatures as indicated in the frames. The lines represent fits using the Havriliak-Negami formula, Eq. (1), as described in the text.

To deduce quantitative information about the time scale of the observed relaxational phenomena, an analysis of the frequency-dependent dielectric data is most appropriate. Figure 4 shows the dielectric-constant spectra for all three samples, measured at a series of temperatures. Again, we focus here on the real part $\varepsilon'$ because in $\varepsilon''$ the dipolar relaxation peak is partly superimposed by the conductivity contributions [36]. In Appendix C, additional plots of $\varepsilon''(\nu)$ are shown for the same temperatures. For all samples, the $\varepsilon'$ spectra of Fig. 4 reveal a steplike decrease with increasing frequency, typical for dipolar relaxation processes [26,28]. The observed shift of this feature to lower frequencies with decreasing temperature evidences the continuous slowing down of the associated hydrogen dynamics within the double-well potential under cooling. As already seen in Figs. 1 and 2, the dipolar strength $\Delta\varepsilon$, corresponding to the step height, exhibits qualitatively different temperature characteristics for the single- and polycrystalline samples. Moreover, Fig. 4 reveals different values of the high-frequency dielectric constant $\varepsilon_\infty$ for the three investigated samples. This may be ascribed to different stray capacitances arising from the strongly different sample geometries (see section II). We did not correct for these effects as they do not affect the analysis of the dipole dynamics.

Via the Onsager equation [37], the size of the fluctuating dipolar moment $\mu$, leading to the observed relaxation steps, can be estimated from $\varepsilon_\infty$ and the measured static dielectric constant $\varepsilon_s$ [38]. As the Onsager equation describes single-dipole motions, we performed this calculation for the single crystal, which does not exhibit ferroelectric correlations and, in addition, has a reasonable value of $\varepsilon_\infty \approx 8$. For 200 K, we arrive at $\mu \approx 0.62$. This is of comparable order as the calculated dipole moment of 0.29 - 0.53 D from Ref. [17].

To fit the spectra of Fig. 4, we employed the Havriliak-Negami equation [39], a generalized form of a Debye relaxation:

$$\varepsilon^*(\nu) = \varepsilon_\infty + \frac{\Delta\varepsilon}{[1+(i2\pi\nu\tau)^{1-\alpha}]^\beta} \qquad (1).$$

Here, $\varepsilon^*(\nu)$ is the complex dielectric constant. The quantity $\Delta\varepsilon = \varepsilon_s - \varepsilon_\infty$ represents the dipolar strength of the relaxational process, governed by the off-centering of the proton (deuteron) in the present case. Furthermore, $\tau$ is the relaxation time characterizing the dynamics of the dipolar motion. The parameters $\alpha \leq 1$ and $\beta \leq 1$ determine the broadening of the predicted loss peaks and steps in $\varepsilon'$, compared to the Debye case. The latter is recovered in Eq. (1) for $\alpha = 0$ and $\beta = 1$. Deviations from the Debye case usually signal a distribution of relaxation times, typical for glasslike systems [40,41].

Representative results of the fits using Eq. (1), simultaneously performed for $\varepsilon'$ and $\varepsilon''$, are illustrated as solid lines in Fig. 4 and in the corresponding $\varepsilon''$ spectra shown in Appendix C. At most of the temperatures investigated, the Havriliak-Negami function provides a reasonable description of the dipolar relaxation process in protonated and deuterated samples in single-crystalline and in ceramic form. For the deuterated compound, small deviations at high frequencies and low temperatures indicate an additional contribution of unknown origin, which may be due to ac-conductivity. This was ignored in the fits, which did not hamper the reliable determination of the relaxation times. From these fits, we can derive the parameters describing broadening and asymmetry, as well as the relaxation times with reasonable accuracy. Notably, the width parameters at all temperatures significantly deviate from the Debye behavior, i.e., $\alpha \neq 0$ and/or $\beta \neq 1$. This points to a distribution of relaxation times, as also commonly found for glass-forming matter [28,40,41].

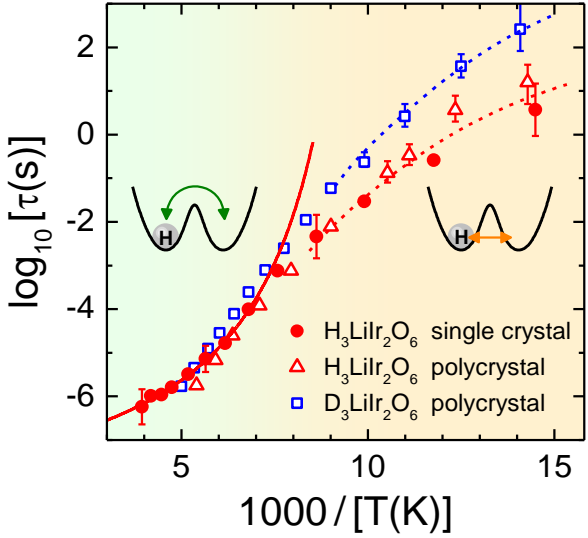

FIG. 5. Arrhenius plot of the relaxation times of the investigated compounds as obtained from fits of the dielectric spectra as discussed in the text. The relaxation times were determined for single- (filled circles) and polycrystalline $H_3LiIr_2O_6$ (empty triangles) as well as for deuterated $D_3LiIr_2O_6$ (empty squares). The solid line represents a fit of the single-crystal data at high temperatures by the VFT law, Eq. (2). The dashed lines are drawn to guide the eye. As discussed in the text, the high-temperature behavior essentially involves thermally activated dynamics, while at low temperatures tunneling processes gain increasing importance. The corresponding motions of the hydrogens within the double well potentials of the hydrogen bonds are schematically indicated in the two insets.

The most relevant parameter in the context of the present work is the relaxation time. Its temperature dependence is shown in Fig. 5 using an Arrhenius representation. At elevated temperatures, a striking super-Arrhenius behavior of $\tau(T)$ is found, which becomes most obvious for the single-crystalline sample. Similar deviations from thermally activated dynamics are commonly found for glass-forming materials and believed to signify increasing cooperativity of the relaxing dipolar entities under cooling [28,42,43]. There the temperature dependence of the dipolar relaxation times often follows the Vogel-Fulcher-Tammann (VFT) equation [44,45,46,47],

$$\tau = \tau_0 \, exp\left(\frac{DT_{VF}}{T - T_{VF}}\right) \quad (2).$$

Here $\tau_0$ is an inverse attempt frequency, $D$ the so-called strength parameter, quantifying the degree of deviation from Arrhenius behavior [47], and $T_{VF}$ is the Vogel-Fulcher temperature, where the extrapolated relaxation time would diverge. This equation describes a super-Arrhenius behavior with a divergence temperature $T_{VF}$ that is often assumed to indicate some kind of hidden phase transition [42,43]. The latter never can be observed because, under cooling, the system falls out of thermodynamic equilibrium at significantly higher temperature, namely at the freezing or glass-transition temperature $T_g$ [42]. The VFT equation is the canonical expression to describe glassy freezing in various types of disordered matter [28,42,47].

The present parameterization of the high-temperature $\tau(T)$ data in terms of the VFT law yields $T_{VF} \approx 92$ K, $D \approx 4.4$, and $\tau_0 \approx 50$ ns (solid line in Fig. 5) [48]. $\tau_0$ seems unusually large as it should correspond to a microscopic time and be of order ps. Applying the usual criterion $\tau(T_g) \approx 100$ s, we arrive at $T_g \approx 110$ K for the hypothetical glass-transition temperature of the hydrogen motion within $H_3LiIr_2O_6$. However, Fig. 5 reveals that, already below about 130 K ($1000/T > 7.7$ K$^{-1}$), $\tau(T)$ deviates from the VFT curve and crosses over into a significantly weaker temperature dependence for all samples. This can be explained by an increasing importance of tunneling at low temperatures, which leads to significantly faster relaxation compared to the extrapolated VFT behavior. While the VFT law in glass physics in principle is based on thermally activated processes with additional increasing cooperativity when approaching the glass transition [28,42,43], the tunneling probability should essentially be temperature independent. This leads to weaker temperature dependence of the relaxation time, when under cooling thermal activation becomes increasingly unlikely. Tunneling in these systems is supported by the fact that the relaxation times within the tunneling regime of the deuterated sample are by about a factor of 10 slower than those of the protonated compounds, as is expected from the mass dependence of quantum-mechanical tunneling. As mentioned above, tunneling of the protons within the double-well potentials of the hydrogen bonds in $H_3LiIr_2O_6$ was also considered in two recent theoretical works [16,17].

## IV. SUMMARY AND CONCLUSIONS

In summary, we have performed detailed dielectric-spectroscopy measurements of $H_3LiIr_2O_6$ and its isostructural deuterated counterpart $D_3LiIr_2O_6$, both having a QSL ground state. For both materials we find a well-pronounced dielectric relaxation process, signifying the previously suggested [16,17] motion of hydrogen ions within the double-well potentials of the hydrogen bonds that couple neighboring $LiIr_2O_6$ honeycomb layers. At elevated temperatures, $T \gtrsim 130$ K, the hydrogen dynamics of both compounds reveals signatures of glassy freezing, documented by a super-Arrhenius behavior of the relaxation time and a broadened non-Debye relaxation [28,40,41,42]. There the hydrogen ions essentially hop via a thermally activated mechanism and $\tau(T)$ shows similar temperature dependence for both compounds. This glassy freezing in the hydrogen bonds also reminds of the super-Arrhenius behavior found for the dipolar degrees of freedom in plastic crystals [49]. However, at lower temperatures the dynamics remains significantly faster than the extrapolated high-temperature behavior, signifying quantum-mechanical tunneling. As expected, in this region the tunneling rate is significantly slower in the deuterated compound.

Our results support theoretical approaches ascribing the QSL state of $H_3LiIr_2O_6$ to a dominant Kitaev exchange due to static or dynamical hydrogen disorder [14,15,16,17]. The detected dipolar relaxation process demonstrates that this disorder in principle is dynamic and that this dynamic nature persists at low temperatures due to tunneling. However, the experimentally observed time scales for tunneling are significantly slower than theoretically estimated [16]. At $T < 100$ K, the hydrogen ions fluctuate in the range of 1 to 100 s (Fig. 5). Compared to magnetically relevant time scales, this corresponds to effectively static hydrogen disorder as considered in Refs. [14,15] and schematically indicated in Fig. 1. One may speculate that the relatively slow hydrogen fluctuation rates at low temperatures are caused by cooperative hydrogen motions. As has been proposed by Knolle *et al*. [15], random positions of $H^+$ or $D^+$ ions can generate local variations of the crystal field on the nearby oxygen ions and thus can introduce local distortions of the oxygen octahedra. The latter in turn modify the magnetic coupling between the iridium ions. The resulting random Kitaev-like bonds as proposed in Ref. [15], not only account for the QSL state but also for the divergent specific heat and the non-vanishing spin-lattice relaxation rate at low temperatures reported in [10].

Among the different explanations of the QSL state in $H_3LiIr_2O_6$ mentioned in the introduction, the scenario of an increased importance of the Kitaev exchange due to hydrogen disorder obviously is well supported by the findings of the present work. However, we cannot completely rule out the possible alternative of disorder-induced spin defects (caused by the detected hydrogen disorder), which may lead to the observed spin-liquid state without having to invoke any Kitaev physics [21]. Such a mechanism was recently also proposed for the suggested QSL phase in the organic charge-transfer salt $\kappa$-$(ET)_2Cu_2(CN)_3$ [50] whose dielectric response, interestingly, reveals very similar relaxor-like behavior as the polycrystalline $H_3LiIr_2O_6$ and $D_3LiIr_2O_6$ samples (Fig. 3) [51]. In any case, the presence of effectively static hydrogen disorder in $H_3LiIr_2O_6$, evidenced by the present work, should play an important role for the formation of a QSL state in this system. Obviously, this material is both a QSL and a quantum paraelectric.

## ACKNOWLEDGMENTS

We gratefully acknowledge stimulating discussions with Philipp Gegenwart. This research was partly funded by the Deutsche Forschungsgemeinschaft via the Transregional Research Collaboration TRR 80: From Electronic Correlations to Functionality.

## APPENDIX A: NUCLEAR MAGNETIC RESONANCE

To document the similar magnetic behavior of single crystals and powder samples, Fig. 6 displays nuclear magnetic resonance (NMR) spectra obtained for single-crystalline samples. The experiment was carried out applying the same methods as for the oriented-powder NMR described in Ref. [10]. The temperature dependence of the Knight shift (the peak positions in Fig. 6) for fields parallel or perpendicular to the honeycomb plane behaves very similar as observed for the oriented powder. No significant broadening of the spectra is seen down to the lowest temperature of 2 K. As already reported in the previous powder study, a tiny increase of the half width at half maximum, 0.04 % in the Knight shift, corresponds to only 0.002 $\mu_B$ per Ir atom, due to impurity or defects. Considering the mentioned correspondence between the single crystal and powder samples, we conclude on a quantitatively similar QSL ground state.

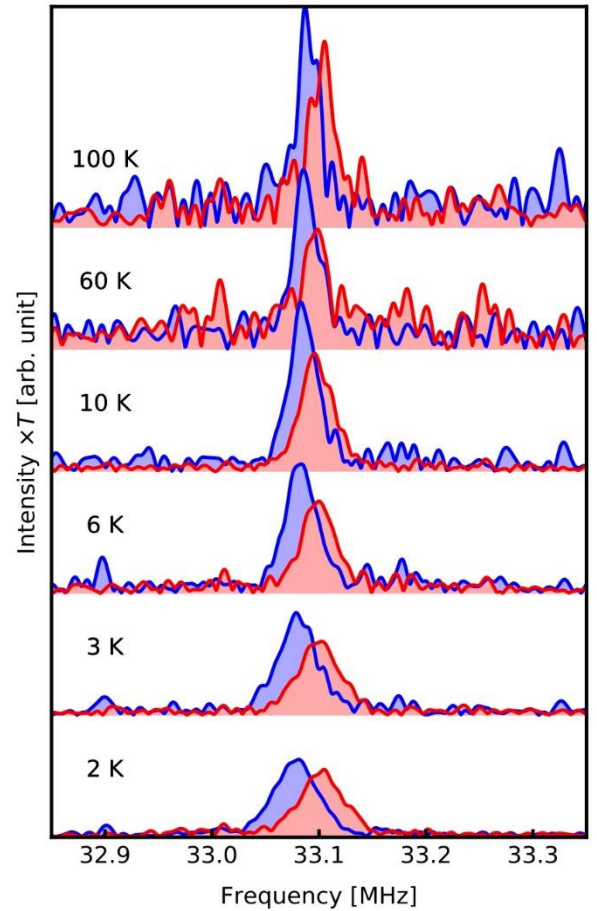

FIG. 6. $^7$Li-NMR spectra taken on single-crystalline $H_3LiIr_2O_6$ at 2 T with the external magnetic field applied parallel (red) or perpendicular (blue) to the honeycomb plane.

## APPENDIX B: ELECTRICAL RESISTIVITY

Figure 7 shows the temperature dependence of the dc resistivity $\rho_{dc}$ in $H_3LiIr_2O_6$ and $D_3LiIr_2O_6$. These data correspond to the inverse of the conductivity $\sigma'$, measured at the lowest probing frequency in the mHz regime as exemplified for the single-crystalline compound in Fig. 2(b). Hence, these results correspond to resistivity experiments in two-point configuration. The data shown in Fig. 7 are restricted to temperatures above the onset of dominating non-dc contributions, which are revealed by a crossover to a significantly weaker temperature dependence of $\sigma'(T)$ at low temperatures [observed, e.g., below about 50 K for the single-crystalline sample, cf. Fig. 2(b)]. The conductivities have been determined for the protonated compound in single-crystalline and ceramic samples and for the deuterated compound in a ceramic sample only. For polycrystalline $H_3LiIr_2O_6$, our results obtained in two-point configuration using dielectric measuring techniques nicely match those measured by conventional dc methods in four-point configuration, as reported in Ref. [10] and shown by the crosses in Fig. 7.

Figure 7(a) represents the data in an Arrhenius representation, plotting the logarithm of the resistivity $\rho_{dc} = 1/\sigma_{dc}$ vs. the inverse temperature. The strong curvature of these plots for all compounds investigated documents that for all samples charge transport cannot be described by simple thermally activated behavior. However, for disordered semiconductors, Mott [30] has derived his time-honored variable range hopping (VRH) law:

$$\rho_{dc} = \rho_0 \exp[(T_0/T)^{1/4}] \qquad (3)$$

Here $\rho_0$ is a prefactor and $T_0$ represents a characteristic energy (in terms of temperature) depending on the third power of the inverse of the localization length and on the density of states. The exponent ¼ signals three-dimensional hopping. This VRH law has been derived under the assumption that the hopping of localized charge carriers depends on their separation in space and in energy. In Fig. 7(b) we plot the logarithm of the resistivities vs. $T^{-1/4}$ and, indeed, find a nearly linear behavior in an extended temperature regime, at least at higher temperatures. At low temperatures, ac contributions may play an increasing role, leading to a somewhat weaker temperature dependence.

Thus, Mott's VRH mechanism in a three-dimensional system seems to be the adequate description of charge transport in these materials. The characteristic temperatures $T_0$ are $5.9 \times 10^8$ K for single crystalline and $8.5 \times 10^8$ K for polycrystalline $H_3LiIr_2O_6$. For the deuterated ceramic sample we find $T_0 = 1.6 \times 10^9$ K. These values are comparable to those reported for various amorphous semiconductors or disordered correlated materials, e.g., the colossal resistive manganates [31,32]. The characteristic temperatures are of similar order for the protonated single-crystalline and ceramic compounds, documenting that the localization length and the electronic density of states are comparable. The values are somewhat larger for the deuterated compound. It is unclear whether this increase of $T_0$ results from changes in the localization length or from an increase in the electronic density of states.

Please note that these experiments have been performed using the geometry of the dielectric experiments. Hence, in the single crystal the resistivity was measured primarily perpendicular to the planes along the crystallographic $c$ direction. In the ceramic samples in-plane and out-of-plane resistivities are averaged out. Figure 7 reveals a significantly higher resistivity for single-crystalline $H_3LiIr_2O_6$ compared to the ceramic sample. Thus, the electronic hopping transport along the $c$ direction perpendicular to the honeycomb layers obviously is significantly reduced compared to in-plane hopping processes.

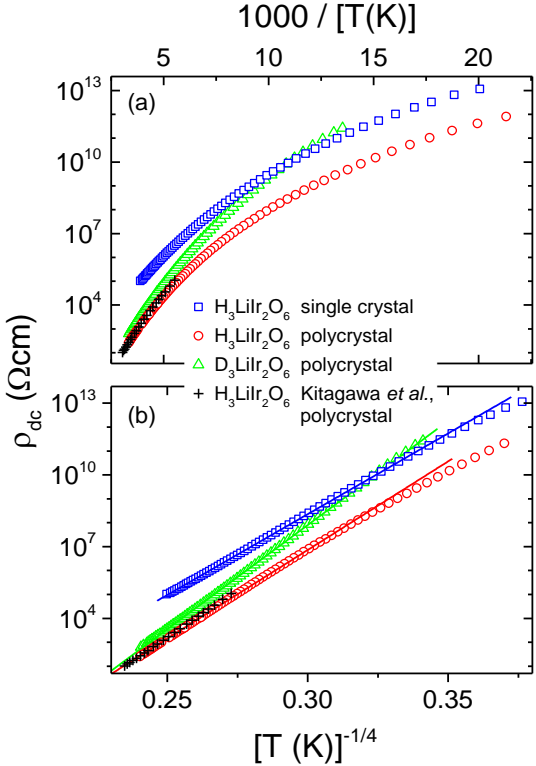

FIG. 7. Temperature dependence of the electrical dc resistivity in ceramic and single-crystalline $H_3LiIr_2O_6$ and $D_3LiIr_2O_6$ as indicated in the figure legend. (a) Arrhenius representation and (b) VRH representation, where the logarithm of the resistivity is plotted vs. $T^{-1/4}$. The crosses indicate the resistivity of $H_3LiIr_2O_6$ as published by Kitagawa *et al*. [10]. The solid lines represent fits using Eq. (3) in the high-temperature regime as described in the text.

## APPENDIX C: DIELECTRIC-LOSS SPECTRA

Figure 8 shows spectra of the dielectric loss for all three samples as measured at various frequencies. The corresponding $\varepsilon'$ spectra are presented in Fig. 4. At low frequencies, $\varepsilon''(\nu)$ is dominated by the dc-conductivity contribution. Due to the relation $\varepsilon'' \propto \sigma'/\nu$, it shows up as a $1/\nu$ divergence of $\varepsilon''(\nu)$ towards low frequencies. The dipolar relaxation should lead to a loss peak, which, however, is partly superimposed by the conductivity contributions. This peak is only revealed by its right flank, leading to a weaker frequency dependence of $\varepsilon''$ at high frequencies [cf. dashed line in Fig. 8(a) indicating the unobscured loss peak for 116 K]. The lines in Fig. 8 are fits using Eq. (1) with an additional contribution $\varepsilon''_{dc} = \sigma_{dc}/(2\pi\nu\varepsilon_0)$, where $\varepsilon_0$ is the permittivity of vacuum. The fits were simultaneously performed for $\varepsilon'$ and $\varepsilon''$. It is clear that significant information on the relaxation time is only obtained from the $\varepsilon'$ spectra.

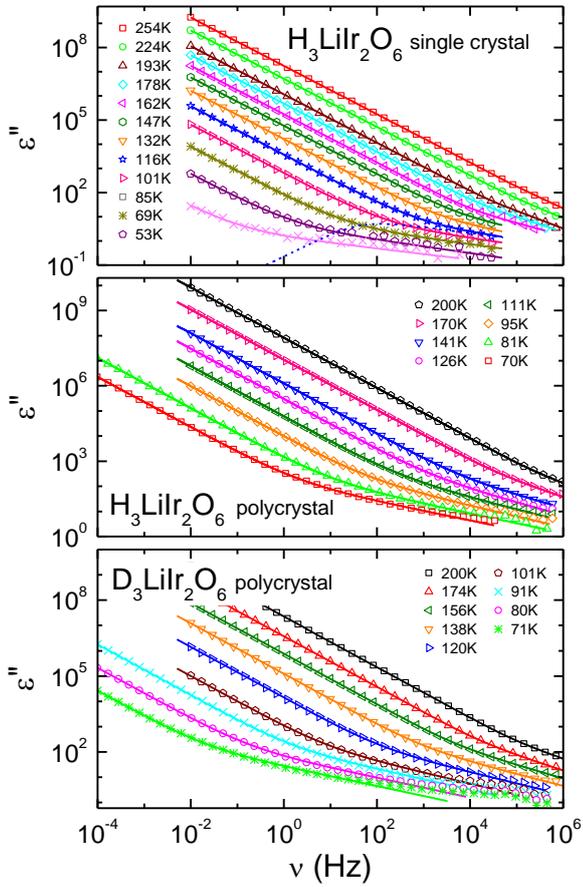

FIG. 8. Frequency dependence of the dielectric loss $\varepsilon''$ for single-crystalline (a) and polycrystalline $H_3LiIr_2O_6$ (b), as well as for deuterated polycrystalline $D_3LiIr_2O_6$ (c). Data are shown for selected temperatures as indicated in the frames. The lines represent fits using the Havriliak-Negami formula, Eq. (1) with an additional $1/\nu$ contribution arising from the dc conductivity, as described in the text. As an example, the dashed line in (a) shows the unobscured relaxation peak for 116 K.

[1] L. Balents, *Spin liquids in frustrated magnets*, Nature **464**, 199 (2010).
[2] J. Knolle and R. Moessner, *A field guide to spin liquids*, Annu. Rev. Condens. Matter. Phys. **10**, 451 (2019).
[3] A. Kitaev, *Anyons in an exactly solved model and beyond*, Ann. Phys. **321**, 2 (2006).
[4] G. Jackeli and G. Khaliullin, *Mott insulators in the strong spin-orbit coupling limit: From Heisenberg to a quantum compass and Kitaev models*, Phys. Rev. Lett. **102**, 017205 (2009).
[5] Y. Singh and P. Gegenwart, *Antiferromagnetic Mott insulating state in single crystals of the honeycomb lattice material $Na_2IrO_3$*, Phys. Rev. B **82**, 064412 (2010).
[6] Y. Singh, S. Manni, J. Reuther, T. Berlijn, R. Thomale, W. Ku, S. Trebst, and P. Gegenwart, *Relevance of the Heisenberg-Kitaev model for the honeycomb lattice iridates $A_2IrO_3$*, Phys. Rev. Lett. **108**, 127203 (2012).
[7] K. W. Plumb, J. P. Clancy, L. J. Sandilands, V. V. Shankar, Y. F. Hu, K. S. Burch, H.-Y. Kee, and Y.-J. Kim, α-RuCl3: a spin-orbit assisted Mott insulator on a honeycomb lattice, Phys. Rev. B **90**, 041112(R) (2014).
[8] S. M. Winter, A. A. Tsirlin, M. Daghofer, J. van den Brink, Y. Singh, P. Gegenwart, and R. Valenti, *Models and materials for generalized Kitaev magnetism*, J. Phys.: Condens. Matter **29**, 493002 (2017).
[9] H. Takagi, T. Takayama, G. Jackeli, G. Khaliullin, and S. E. Nagler, *Concept and realization of Kitaev quantum spin liquids*, Nat. Rev. Phys. **1**, 264 (2019).
[10] K. Kitagawa, T. Takayama, Y. Matsumoto, A. Kato, R. Takano, Y. Kishimoto, S. Bette, R. Dinnebier, G. Jackeli, and H. Takagi, *A spin-orbital-entangled quantum liquid on a honeycomb lattice*, Nature **554**, 341 (2018).
[11] S. Bette, T. Takayama, K. Kitagawa, R. Takano, H. Takagia, and R. E. Dinnebier, *Solution of the heavily stacking faulted crystal structure of the honeycomb iridate $H_3LiIr_2O_6$*, Dalton Trans. **46**, 15216 (2017).
[12] S. Pei, L.-L. Huang, G. Li, X. Chen, B. Xi, X. W. Wang, Y. Shi, D. Yu, C. Liu, L. Wang, F. Ye, M. Huang, and J.-W. Mei, *Magnetic Raman continuum in single crystalline $H_3LiIr_2O_6$*, arXiv: 1906.03601
[13] T. Takayama, *Exotic honeycomb magnets with strong spin-orbit coupling*, in *Proceedings of the APS March Meeting, 2018* (Bulletin of the American Physical Society, 2018).
[14] R. Yadav, R. Ray, M. S. Eldeeb, S. Nishimoto, L. Hozoi, and J. van den Brink, *Strong effect of hydrogen order on magnetic Kitaev interactions in $H_3LiIr_2O_6$*, Phys. Rev. Lett. **121**, 197203 (2018).
[15] J. Knolle, R. Moessner, and N. B. Perkins, *Bond-disordered spin liquid and the honeycomb iridate $H_3LiIr_2O_6$: Abundant low-energy density of states from random Majorana hopping*, Phys. Rev. Lett. **122**, 047202 (2019).
[16] Y. Li, S. M. Winter, and R. Valenti, *Role of hydrogen in the spin-orbital entangled quantum liquid candidate $H_3LiIr_2O_6$*, Phys. Rev. Lett. **121**, 247202 (2018).
[17] S. Wang, L. Zhang, and F. Wang, *Possible quantum paraelectric state in the Kitaev spin liquid candidate $H_3LiIr_2O_6$*, arXiv: 1807.03092.
[18] K. Slagle, W. Choi, L. E. Chern, and Y. B. Kim, *Theory of quantum spin liquid in the hydrogen intercalated honeycomb iridate $H_3LiIr_2O_6$*, Phys. Rev. B **97**, 115159 (2018).
[19] R. Yadav, M. S. Eldeeb, R. Ray, S. Aswartham, M. I. Sturza, S. Nishimoto, J. van den Brink, and L. Hozoi, *Engineering Kitaev exchange in stacked iridate layers: impact of inter-layer species on in-plane magnetism*, Chem. Sci. **10**, 1866 (2019).
[20] For low energies, the specific-heat and NMR results for $H_3LiIr_2O_6$ suggest an abundant density of states, signified, e.g., by $C/T \propto T^{-1/2}$ [10], which is at odds with the vanishing specific heat ($C/T \propto T$) expected for a Kitaev QSL at low $T$.
[21] I. Kimchi, J. P. Sheckelton, T. M. McQueen, and P. A. Lee, *Scaling and data collapse from local moments in frustrated disordered quantum spin systems*, Nature Commun. **9**, 4367 (2018).
[22] E. Courtens, *Vogel-Fulcher Scaling of the Susceptibility in a Mixed-Crystal Proton Glass*, Phys. Rev. Lett. **52**, 69 (1983).
[23] S. L. Hutton, I. Fehst, R. Böhmer, M. Braune, B. Mertz, P. Lunkenheimer, and A. Loidl, *Proton glass behavior and hopping conductivity in solid solutions of antiferroelectric betaine phosphate and ferroelectric betaine phosphite*, Phys. Rev. Lett. **66**, 1990 (1991).
[24] U. T. Höchli, K. Knorr, and A. Loidl, *Orientational Glasses*, Advances in Physics **39**, 405 (1990).
[25] U. Schneider, P. Lunkenheimer, A. Pimenov, R. Brand, and A. Loidl, *Wide Range Dielectric Spectroscopy on Glass-Forming Materials: An Experimental Overview*, Ferroelectrics **249**, 89 (2001).
[26] A. Schönhals and F. Kremer, *Analysis of Dielectric Spectra*, in *Broadband Dielectric Spectroscopy*, edited by F. Kremer and A. Schönhals (Springer, Berlin, 2002), p. 59
[27] P. Lunkenheimer and A. Loidl, *Dielectric spectroscopy on organic charge-transfer salts*, J. Phys.: Condens. Matter **27**, 373001 (2015).
[28] P. Lunkenheimer and A. Loidl, *Glassy dynamics: From millihertz to terahertz*, in: *The scaling of relaxation processes*, edited by F. Kremer and A. Loidl (Springer, Cham, 2018), p. 23.
[29] S. Emmert, M. Wolf, R. Gulich, S. Krohns, S. Kastner, P. Lunkenheimer, and A. Loidl, *Electrode polarization effects in broadband dielectric spectroscopy*, Eur. Phys. J. B **83**, 157 (2011).
[30] N. F. Mott, *Charge transport in non-crystalline semiconductors*, Festkörperprobleme **9**, 22 (1969).
[31] A. Seeger, P. Lunkenheimer, J. Hemberger, A. A. Mukhin, V. Yu. Ivanov, A. M. Balbashov, and A. Loidl, *Charge carrier localization in $La_{1-x}Sr_xMnO_3$ investigated by ac conductivity measurements*, J. Phys.: Condens. Matter **11**, 3273 (1999).
[32] M. Paraskevopoulos, F. Mayr, J. Hemberger, A. Loidl, R. Heichele, D. Maurer, V. Müller, A. A. Mukhin, and A. M. Balbashov, *Magnetic properties and the phase diagram of $La_{1-x}Sr_xMnO_3$ for $x < 0.2$*, J. Phys.: Condens. Matter **12**, 3993 (2000).
[33] L. E. Cross, *Relaxor ferroelectrics*, Ferroelectrics **76**, 241 (1987).
[34] G. A. Samara, *The relaxational properties of compositionally disordered $ABO_3$ perovskites*, J. Phys. Condens. Matt. **15**, R367 (2003).
[35] D. Viehland, S. J. Jang, L. E. Cross, and M. Wuttig, Freezing of the polarization fluctuations in lead magnesium niobate relaxors, J. Appl. Phys. **68**, 2916 (1990).
[36] The plots in Fig. 4 are restricted to the frequency/temperature regions where the observed relaxation process dominates the measured spectra.
[37] C. J. F. Böttcher and P. Bordewijk, *Theory of electric polarization*, Vol. I (Elsevier, Amsterdam, 1973).

[38] $\varepsilon_\infty$, which usually is only weakly temperature dependent, is best determined at low temperatures, where the high-frequency plateau of the $\varepsilon'$ spectra is well developed. $\varepsilon_s$ can be deduced from the low-frequency plateau. Reading off these values in Fig. 4(a) leads to $\varepsilon_\infty \approx 8$ and $\varepsilon_s \approx 31$ at 200 K. The number density of dipoles included in the Onsager equation, was calculated assuming six H atoms per unit cell and a cell volume of ~222 Å$^3$ [11].

[39] S. Havriliak and S. Negami, *A complex plane analysis of α-dispersions in some polymer systems*, J. Polym. Sci. C **14**, 99 (1966).

[40] H. Sillescu, *Heterogeneity at the glass transition: a review*, J. Non-Cryst. Solids **243**, 81 (1999).

[41] M. D. Ediger, Spatially *Heterogeneous Dynamics in Supercooled Liquids*, Annu. Rev. Phys. Chem. **51**, 99 (2000).

[42] M. D. Ediger, C. A. Angell, and S. R. Nagel, *Supercooled Liquids and Glasses*, J. Phys. Chem. **100**, 13200 (1996).

[43] S. Albert, Th. Bauer, M. Michl, G. Biroli, J.-P. Bouchaud, A. Loidl, P. Lunkenheimer, R. Tourbot, C. Wiertel-Gasquet, and F. Ladieu, *Fifth-order susceptibility unveils growth of thermodynamic amorphous order in glass-formers*, Science **352**, 1308 (2016).

[44] H. Vogel, *Das Temperaturabhängigkeitsgesetz der Viskosität von Flüssigkeiten*, Phys. Z. **22**, 645 (1921).

[45] G. S. Fulcher, *Analysis of recent measurements of the viscosity of glasses*, J. Am. Ceram. Soc. **8**, 339 (1925).

[46] G. Tammann and W. Hesse, *Die Abhängigkeit der Viskosität von der Temperatur bei unterkühlten Flüssigkeiten*, Z. Anorg. Allg. Chem. **156**, 245 (1926).

[47] C. A. Angell, *Strong and Fragile Liquids*, in *Relaxations in Complex Systems*, edited by K. L. Ngai and G. B. Wright (NRL, Washington, DC, 1985), p 1.

[48] We have fitted the single-crystal data, which extend to higher temperatures than the data for the other samples.

[49] Th. Bauer, M. Köhler, P. Lunkenheimer, A. Loidl, and C.A. Angell, *Relaxation dynamics and ionic conductivity in a fragile plastic crystal*, J. Chem. Phys. **133**, 144509 (2010).

[50] K. Riedl, R. Valentí, and S. M. Winter, *Critical spin liquid versus valence-bond glass in a triangular-lattice organic antiferromagnet*, Nature Commun. **10**, 2561 (2019).

[51] M. Abdel-Jawad, I. Terasaki, T. Sasaki, N. Yoneyama, N. Kobayashi, Y. Uesu, and C. Hotta, *Anomalous dielectric response in the dimer Mott insulator κ-(BEDT-TTF)$_2$Cu$_2$(CN)$_3$*, Phys. Rev. B **82**, 125119 (2010).